# Optical spectral pulse shaping by combining two oppositely chirped fiber Bragg grating

**Miguel A. Preciado, Víctor García-Muñoz, Miguel A. Muriel**

*ETSI Telecomunicación, Universidad Politécnica de Madrid (UPM), 28040 Madrid, Spain.*
*miguel.preciado@tfo.upm.es, victorgm@tfo.upm.es, muriel@tfo.upm.es*

**Abstract:** In this letter we present a new technique for pulse shaping. The desired pulse is shaped by two apodized chirped fiber Bragg gratings which dispersions are adjusted to be cancelled. This technique exploits the well-known property of linearly-chirped gratings, in which the apodization (amplitude) grating profile maps its spectral response. This technique presents inherent advantages of chirped fiber Bragg gratings and direct design in frequency domain.

**References and Links**

1. A. M. Weiner, "Femtosecond optical pulse shaping and processing," Prog. Quant. Electron. **19**, 161–235 (1995).
2. T. Kurokawa, H. Tsuda, K. Okamoto, K. Naganuma, H. Takenouchi, Y. Inoue, and M. Ishii, "Time-space-conversion optical signal processing using arrayed-waveguide grating," Electron. Lett. **33**, 1890-1891 (1997).
3. P. Emplit, M. Haelterman, R. Kashyap, and M. De Lathouwer, "Fiber Bragg grating for optical dark soliton generation," IEEE Photon. Technol. Lett. **9**, 1122–1124 (1997).
4. G. Curatu, S. LaRochelle, C. Pare and P-A. Belanger, "Antisymmetric pulse generation using phase-shifted fibre Bragg grating," Electron. Lett. **38**, 307-309 (2002).
5. P. Petropoulos, M. Ibsen, A.D. Ellis, D.J. Richardson, "Rectangular pulse generation based on pulse reshaping using a superstructured fiber Bragg grating," J. Lightwave Technol. **19**, 746-752 (2001).
6. F. Parmigiani, P. Petropoulos, M. Ibsen, D.J. Richardson "All-optical pulse reshaping and retiming systems incorporating pulse shaping fiber Bragg grating," J. Lightwave Technol. **24**, 357-364 (2006).
7. R. Feced, M. N. Zervas, and M. A. Muriel, "An efficient inverse scattering algorithm for the design of non uniform fiber Bragg gratings,'' IEEE J. of Quant. Electron. **35**, 1105–1115 (1999).
8. L. Poladian, "Simple grating synthesis algorithm," Opt. Lett. **25**, 787–789 (2000).
9. S. Longhi, M. Marano, P. Laporta, O. Svelto, "Propagation, manipulation, and control of picosecond optical pulses at 1.5 µm in fiber Bragg gratings, J. Opt. Soc. Am. B **19**, 2742-2757 (2002).
10. J. Azaña , and L. R. Chen, "Synthesis of temporal optical waveforms by fiber Bragg gratings: a new approach based on space-to-frequency-to-time mapping", J. Opt. Soc. Am. B **19**, 2758-2769 (2002).
11. J. Azaña and M. A. Muriel, "Technique for multiplying the repetition rates of periodic trains of pulses by means of a temporal self-imaging effect in chirped fiber gratings," Opt. Lett. **24**, 1672–1674 (2000).
12. F. Ouellette, "Dispersion cancellation using linearly chirped Bragg grating filters in optical waveguides," Opt. Lett. 12, 847- (1987)
13. A. G. Jepsen, A. E. Johnson, E. S. Maniloff, T. W. Mossberg, M. J. Munroe, and J. N. Sweetser, "Fibre Bragg grating based spectral encoder/decoder for lightwave CDMA," Electron. Lett. **35**, 1096-1097 (1999).
14. J. Azaña and M. A. Muriel, ''Real-time optical spectrum analysis based on the time-space duality in chirped fiber gratings,'' IEEE J. of Quant. Electron. **36**, 517–527 (2000).
15. B. Bovard, "Derivation of a matrix describing a rugate dielectric thin film," Appl. Opt. **27**, 1998–2004 (1988).
16. L. Poladian, "Understanding profile-induced group-delay ripple in Bragg gratings," Appl. Opt. **39**, 1920–1923 (2000).

## 1. Introduction

Optical pulse shaping and manipulation are critical features for ultrafast optics, playing a central role in the area of optical communication. For many years, various all-optical techniques has been deployed to deal with this task. In the framework of free-space optics, different pulse shaping techniques based on spatial masking have been proposed [1]. In these works, the authors report impressive results, but the inherent free-space optics limitations restrict this approach. Alternative techniques based on Arrayed Waveguide Gratings [2] or fiber Bragg gratings (FBGs) have been proposed.

In this paper, we focus our attention on optical pulse shaping using FBGs, which have been used as frequency-filtering stages in some pulse-shaping applications [3-6]. Under Born approximation the design process is widely simplified because the corresponding reflection temporal impulse response is directly related to the apodization profile. On the other hand, Born approximation includes a limitation of the strength of the grating and the length of the output pulse. These limitations can be avoided by using grating synthesis algorithms, such as inverse-scattering and layer-peeling techniques [7,8], but these algorithms do not always ensure the feasibility of the resulting grating profile. Chirped FBGs have also been used in optical pulse shaping [9], [10] with high reflectivity without the limitations previously commented. In [9] a technique for repetition-rate multiplication and pulse reshaping by use of an apodized chirped FBG is proposed. Apodization profile is designed for spectral shaping, and dispersion is designed to cause Talbot effect [11], so that this approach only works at a discrete set of pulse train rates. In [10] the apodization profile is directly related to the reflective temporal impulse response amplitude, but a quadratic phase term is added, which limits its use to insensible phase systems.

In this letter a technique based on chirped FBGs is presented. As it can be seen in Fig. 1, the system includes two chirped FBGs connected by an optical circulator. Under high dispersion regime, the apodization profile of an FBG is directly related to its spectral response. The first grating provides the spectral amplitude for pulse shaping, and the second one is a dispersion compensator [12], which cancels the dispersion introduced by the first grating. A similar architecture, two oppositely chirped FBGs, has been previously proposed in CDMA [13]. In this approach, phase-shifts are introduced in the spectra of the output signal to generate spectral-phase-encoded bit, but no considerations about the amplitude of the apodization profile or the spectral amplitude are made.

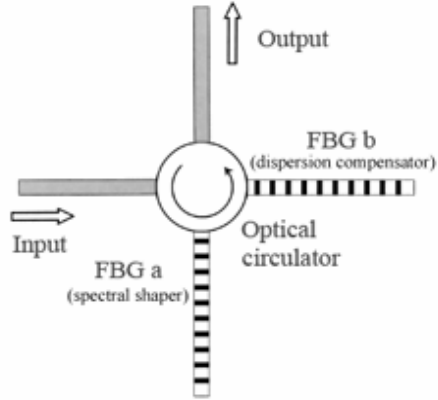

Fig. 1. Architecture of the system. Input signal is processed by two oppositely chirped FBGs, which are connected by an optical circulator.

## 2. Theory

At this point, we introduce the theoretical basis needed to design the system. Suppose a linearly chirped FGB with reflected spectral response $H_R(\omega) = \sqrt{R(\omega)} \cdot e^{j\phi_R(\omega)}$, where $\omega$ is the angular frequency, $R(\omega)$ is the reflectivity, and $\phi_R(\omega)$ is the phase. The refractive index $n(z)$ can be written as:

$$n(z) = n_{av}(z) + \frac{\Delta n_{max}}{2} A(z) \cos[\frac{2\pi}{\Lambda_o} z + \varphi(z)] \qquad (1)$$

where $n_{av}(z)$ represents the average refractive index of the propagation mode, $\Delta n_{max}$ describes the maximum refractive index modulation, $A(z)$ is the normalized apodization function, $\Lambda_o$ is the fundamental period of the grating, $\varphi(z)$ describes the additional phase variation (chirp), and $z \in [0,L]$ is the spatial coordinate over the grating, with $L$ the length of the grating.

In the following we consider a constant average refractive index $n_{av} = n_{eff} + \Delta n_{max}$, where $n_{eff}$ is the effective refractive index of the propagation mode. The additional phase variation can be expressed as $\varphi(z) = (C_k/2)(z - L/2)^2$, where $C_k$ represents the chirp factor, and can be calculated from [14]:

$$C_K = -4n_{av}^2 /(c^2 \ddot{\phi}_R) \tag{2}$$

where $\ddot{\phi}_R = \partial^2 \phi_R(\omega)/\partial \omega^2$ is the first order dispersion coefficient. Besides, the length of the grating $L$ can be obtained from the following expression [14]:

$$L = \left|\ddot{\phi}_R\right| c \Delta \omega_g / (2n_{av}) \tag{3}$$

where $c$ is the light vacuum speed, and $\Delta\omega_g$ is the grating bandwidth. In high dispersion regime, both the temporal and spectral response of an FBG, have been proved to have the same envelope [14]. This high dispersion condition can be expressed as:

$$\left|\Delta t^2 / 8\pi \ddot{\phi}_R\right| \ll 1 \tag{4}$$

where $\Delta t$ is the temporal length of the inverse Fourier transform of the spectral response without the dispersive term, which is approximately equal to the temporal length of the pulse reshaped. From (3) and (4) we can deduce that the shorter temporal length of the pulse, the shorter minimum length of the grating. Notice that the length of the grating is not fixed, but limited by a mimimum, so the length of the grating can be longer that this minimum length.

If condition (4) is met and Born approximation is applicable, both temporal and spectral envelopes reproduces the shape of the apodized function, so we can obtain the apodization profile which corresponds to a desired reflectivity $R_d(\omega)$ [9], that can be written as:

$$\tilde{A}(\omega) = \left[ R_d(\omega) \frac{32 n_{av}^2}{\pi \omega_0^2 \left|\ddot{\phi}_R\right| \Delta n_{max}^2} \right]^{\frac{1}{2}} \tag{5}$$

where $\omega_0$ is central angular frequency and $\tilde{A}(\omega)$ is related to the apodization function as:

$$\tilde{A}(\omega)\bigg|_{\left(\omega = \omega_0 \pm \frac{\Delta \omega_g}{L}(z - L/2)\right)} = A(z) \tag{6}$$

where the sign of $\pm$ is equal to the sign of $C_k$. In the case of high reflectivity an approximate function [15] must be applied over the desired reflectivity $R_d(\omega)$. In particular, here a logarithmic based function is used:

$$\tilde{A}(\omega) = \left[ -\ln\left(1 - R_d(\omega)\right) \frac{32 n_{av}^2}{\pi \omega_0^2 \left|\ddot{\phi}_R\right| \Delta n_{max}^2} \right]^{\frac{1}{2}} \quad (7)$$

### 3. Example and results

As instance we design a system in which gaussian pulses from a short pulse source are reshaped in triangular ones. We assume a carrier frequency ($\omega_0/2\pi$) of 193 THz. Each Gaussian input pulse has an FWHM of 0.7496 ps, and the total desired width for the reshaped triangular pulse is 10 ps. Thus, the spectral function for the input and output pulses are $F_{in}(\omega) \propto \exp\left(-(\omega-\omega_0)^2/\delta_{in}^2\right)$ and $F_{out}(\omega) \propto \text{sinc}^2\left((\omega-\omega_0)/\delta_{out}\right)$ respectively, where $\delta_{in}$=4.443 × 10$^{12}$ and $\delta_{out}$=1.2566 × 10$^{12}$. Notice that $F_{in}(\omega)$ and $F_{out}(\omega)$, as well as all the spectral functions in the following, are described as analytical signals (only defined at $\omega>0$.) We consider a band of interest ($\Delta\omega/2\pi$) of 2 THz centred at $\omega_0$ ($\omega_0$-$\Delta\omega/2 \leq \omega \leq \omega_0$+$\Delta\omega/2$). Fig. 2 shows the schematic diagram. In the following, subscripts 'a' and 'b' refer to first and second FBG respectively.

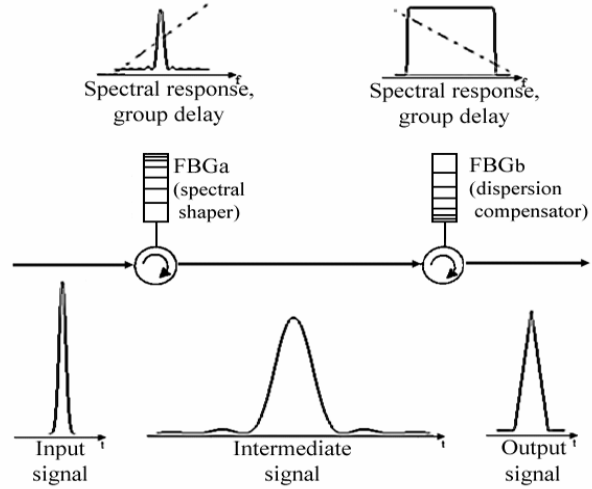

Fig. 2. Schematic diagram. Each FBG with its corresponding spectral response and group delay are showed above. Temporal signals are showed below.

We assume an effective refractive index $n_{eff}$=1.45 for both FBGs. The spectral response of the system meets the following condition:

$$\frac{F_{out}(\omega)}{F_{in}(\omega)} = H_{syst}(\omega) = \left|H_{syst}(\omega)\right| \exp(j\phi_{syst}(\omega)) = $$
$$= \left(R_a(\omega) R_b(\omega)\right)^{1/2} \exp\left(j\left(\phi_{Ra}(\omega) + \phi_{Rb}(\omega)\right)\right) \quad (8)$$

where $H_{syst}(\omega)$ and $\phi_{syst}(\omega)$ are the spectral response and the phase of the system, $R_a(\omega)$, $R_b(\omega)$, $\phi_{Ra}(\omega)$, $\phi_{Rb}(\omega)$ are the reflectivity and phase of both *FBGs*. Thus, we obtain:

$$H_{syst}(\omega) \propto \operatorname{sinc}^2\left((\omega-\omega_0)/\delta_{out}\right)\exp\left((\omega-\omega_0)^2/\delta_{in}^2\right) \quad (9)$$

Relation (8) allows for multiple feasible solutions for $R_a(\omega)$, $R_b(\omega)$. In this approach, we consider that the shape of the reflectivity is influenced by the first FBG solely. This approximation implies that the second FBG presents a flat reflectivity in the band of interest, and we find that:

$$R_a(\omega) = C_a \operatorname{sinc}^4\left(\frac{(\omega-\omega_0)}{\delta_{out}}\right)\exp\left(\frac{2(\omega-\omega_0)^2}{\delta_{in}^2}\right) \quad (10)$$

where $C_a=0.1$ is a constant of design. Using expression (4), we have the dispersion parameter of the first FBG, $\left|\ddot{\phi}_{Ra}\right| \gg 3.979\times 10^{-24} s^2/rad$, where have been used $\Delta t \approx 10$ ps. We choose $\ddot{\phi}_{Ra} = -2.5\times 10^{-22} s^2/rad$. Besides, using Eq. (7) for FBG$_a$ with Eq. (10) at $\omega=\omega_0$ (where $\tilde{A}_a(\omega_0)=1$ is imposed,) we obtain $\Delta n_{max,a}=7.8372 \times 10^{-5}$, $n_{av,a}=1.45008$. Also, we make use of Eq. (2) to calculate the values $C_{Ka} = 3.7434 \times 10^5$ rad/m. From Eq. (3) we obtain $L_a=32.47$ *cm*, where $\Delta\omega_{g,a}=\Delta\omega$ have been assumed. Using Eqs. (6), (7) and (10) we derive:

$$A_a(z) = \left[\frac{1}{\ln(1-C_a)}\ln\left(1-C_a\operatorname{sinc}^4\left(\frac{1}{0.1L_a}\left(z-\frac{L_a}{2}\right)\right)\exp\left(\frac{\left(z-\frac{L_a}{2}\right)^2}{0.0625L_a^2}\right)\right)\right]^{1/2} \quad (11)$$

Besides, the FBG$_b$ must be designed as a dispersion compensator with $\ddot{\phi}_{Rb} = -\ddot{\phi}_{Ra} = 2.5\times 10^{-22} s^2/rad$, and an approximately flat reflectivity in the band of interest.

Fig. 3 shows the output pulse of the system in temporal domain obtained from numerical simulation. Notice that it exhibits the desired triangular shape. In our simulations we assume an ideal spectral response for the dispersion compensator (implemented by FBG$_b$), such that the group-delay ripple [16] under consideration is introduced by FBG$_a$.

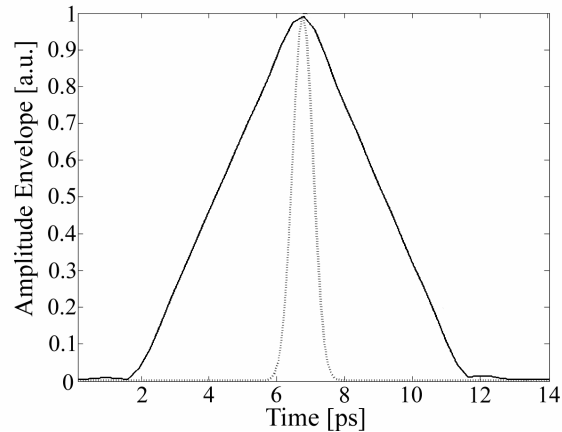

Fig. 3. Amplitude envelope of input pulse (solid line) and output pulse (dashed line.)

## 4. Conclusion

In this paper, we have presented a spectral pulse shaping technique based on a pair of oppositely chirped FBGs. The necessary theoretical basis for designing the FBGs has been developed. Also, we show an example which reshapes Gaussian pulses into triangular ones. Numerical simulation confirms the validity of this approach.

We want to remark that although this architecture has been previously proposed to introduce phase shifts in the spectra of a pulse for CDMA [13], this is the first time, to our knowledge, that it has been used to reshape the spectral amplitude of a pulse.

This approach has the inherent advantages of FBGs over non-FBGs techniques. Although some FBGs pulse shaping techniques (such as Super-structured FBGs) have become extremely sophisticated, two main features can make our novel approach interesting. Firstly, since shaping is made in the spectral domain, it can be very fitting when the desired shape is much simpler in the spectral domain. Secondly, in other techniques, the length of the grating is fixed by the temporal length of the desired reshaped pulse. Thus, there is an unavoidable technological limit when the temporal length is very short, especially if we have a complicated temporal shape. In contrast, in our technique, the length of the grating is limited by a given minimum value, without limitation to present a longer length. Therefore, this novel approach presents less technological restrictions in many cases.

## Ackowledges

This work was supported by the Spanish Ministerio de Educacion y Ciencia under Project "Plan Nacional de I+D+I TEC2004-04754-C03-02".